\documentclass[
 reprint,
 superscriptaddress,
 nofootinbib,
 prx,
 floatfix,
]{revtex4-2}

\usepackage{graphicx}
\usepackage{dcolumn}
\usepackage{bm}
\usepackage{braket}
\usepackage[hidelinks,colorlinks=true]{hyperref}
\usepackage{nicefrac}
\usepackage{stmaryrd}
\usepackage{multirow, makecell}
\usepackage[normalem]{ulem}
\usepackage{tikz}
\usetikzlibrary{quantikz2}

\begin{document}

\title{A Quantum-HPC Hybrid Workflow for Reaction-Center Electronic Dynamics: Application to a Cytochrome P450-Inspired Iron-Complex Model}

\def\quantinuumTokyo{Quantinuum K.K., Otemachi Financial City Grand Cube 3F, 1-9-2 Otemachi, Chiyoda-ku, Tokyo, Japan}
\def\kyotoU{Graduate School of Medicine, Kyoto University, 53 Kawahara-cho, Shogoin Sakyo-ku, Kyoto, Japan}
\def\quantinuumCambridge{Quantinuum, Terrington House, 13-15 Hills Road, Cambridge CB2 1NL, UK}
\def\riken{RIKEN Center for Computational Science, 7-1-26 Minatojima-minamimachi, Chuo-ku, Kobe, Hyogo 650-0047, Japan}

\author{Shintaro~Maekawa}
\email{shintaro.maekawa@quantinuum.com}
\affiliation{\quantinuumTokyo}

\author{Takao~Otsuka}
\affiliation{\kyotoU}

\author{Riku~Masui}
\affiliation{\quantinuumTokyo}

\author{Juan~W.~Pedersen}
\affiliation{\quantinuumTokyo}

\author{David~Mu\~noz Ramo}
\affiliation{\quantinuumCambridge}

\author{Yasushi~Okuno}
\affiliation{\kyotoU}
\affiliation{\riken}

\author{Kentaro~Yamamoto}
\email{kentaro.yamamoto@quantinuum.com}
\affiliation{\quantinuumTokyo}

\date{\today}

\begin{abstract}
We introduce population-transfer dynamics as a practical validation observable for active-space-derived reduced Hamiltonians in multistate reaction-center chemistry. In transition-metal systems, missing couplings or configurations can compromise the reliability of reduced active-space models along a reaction coordinate, even when the overall spectrum is reproduced reasonably well. Using a cytochrome P450-inspired Fe-complex model, we construct a reaction-coordinate-dependent effective Hamiltonian from state-averaged complete active-space self-consistent field (SA-CASSCF) calculations, map it to a quantum-circuit representation suitable for current hardware, and propagate the dynamics from the reactant-side ground state.

At the spectrum level, the reduced Hamiltonian reproduces the SA-CASSCF reference with an RMS deviation of 0.030 eV and a maximum absolute deviation of 0.143 eV. As a dynamics-based diagnostic, the product-manifold population $p_P(t)$, identifies a pronounced near-degeneracy window around $x \approx 0.3$, where state mixing is strongest; specifically, $p_P(t)$ computed from classical exact time evolution reaches 0.488 at $x = 0.3$ after 10 fs, compared with $7.26 \times 10^{-2}$ at $x = 0.2$ and $5.90 \times 10^{-3}$ at $x = 0.0$. To render the workflow executable on current trapped-ion hardware, we quantify the NISQ-relevant trade-off between dynamical fidelity and circuit resources through coupling pruning and first-order Trotterization. A coupling cutoff of $\epsilon = 0.02$ eV reduces the non-zero coupling set from 32 to 7 while preserving the dominant transfer pathways, and a Trotter step number of $M = 30$ provides the best practical operating point among the tested values.

Finally, we demonstrate the end-to-end workflow on Quantinuum’s trapped-ion quantum computer Reimei. The hardware reproduces the key reaction-coordinate trend identified by the classical reduced model, including the pronounced maximum at $x = 0.3$, where the measured product population is 0.42 on hardware and 0.43 on the matched emulator. Overall, this work establishes a dynamics-based diagnostic route for assessing the adequacy of active-space-derived reduced Hamiltonians and shows that chemically interpretable multistate electronic dynamics can already be executed on current trapped-ion hardware under realistic device constraints.
\end{abstract}

\maketitle

\section{Introduction\label{sec:introduction}}

Many chemical reactions in biomolecular systems proceed by exploiting a diverse manifold of electronic states localized at the reaction center. Representative examples include transition-metal enzyme reaction centers such as cytochrome P450 and the photosynthetic oxygen-evolving complex (Mn$_4$CaO$_5$)~\cite{Umena2011}, a strongly-correlated transition-metal complex surrounded by thousands of atoms of the photosystem II (PSII) protein complex~\cite{Yano2014}, which serves as its environment. If such biomolecular systems could be simulated accurately, they would enable clearer mechanistic interpretation of spectroscopy/kinetics and provide design principles for bio-inspired catalysts and selective oxidation chemistry.

In practice, the interpretation of experimental results often relies on reduced models that do not explicitly account for environmental effects~\cite{Yano2014,Meunier2004}. For a rigorous first-principles description~\cite{Szabo1996,Helgaker2000}, however, a hierarchical modeling strategy is essential. Specifically, while the reaction center must be treated with high-precision quantum chemistry, the surrounding environment must be handled using approximations suited for systems with many degrees of freedom. Despite this separation of roles, the combination of strong electronic correlations in the reaction center and the vast configuration space of the surrounding environment remains a major barrier to predictive simulations~\cite{Migliore2014}. The most serious bottleneck is that, even with state-of-the-art high-performance computing (HPC), the computational resources required to evaluate the electronic states of the reaction center can be prohibitively large~\cite{Vogiatzis2017}.

This difficulty is particularly acute in transition-metal enzymes and photochemical reaction systems. In transition-metal systems, nontrivial electronic-structure effects and inherent multiconfigurational character present significant challenges to standard computational approaches~\cite{Vogiatzis2019}. In photochemical and nonadiabatic dynamics, multiple electronic states often interact and become strongly mixed, and their character and couplings may evolve along the reaction coordinate~\cite{Crespo-Otero2018,Lischka2018}. Because nonadiabatic couplings depend on the quality and character of the underlying electronic wave functions, agreement in relative energies alone may not guarantee a consistent description of interstate mixing~\cite{Faraji2018}. Consequently, elucidating the reaction mechanism requires a consistent multistate model of the reaction center that includes interstate couplings and nonadiabatic effects. A practical framework for achieving such a description is multireference electronic-structure theory within a suitably chosen active space~\cite{Stein2016,King2022}.

However, the construction of the active space is inherently path-dependent~\cite{Jeong2020}. Even if the active space is optimal near the reactant region, it may become incomplete in regions where the dominant electronic configurations change, such as near conical intersections. In practice, as the reaction coordinate evolves, ``important configurations'' and ``dominant orbitals'' can interchange; therefore, an active space defined by a single geometric structure may not remain appropriate throughout the entire pathway~\cite{Stein2016,Jeong2020}. Therefore, a framework to verify whether the effective (reduced) model derived from the selected active space is valid across the entire reaction coordinate is essential for ensuring the reliability of mechanistic interpretations~\cite{Stein2016}. In particular, the reduced model should reproduce not only energy but also the interstate mixing that governs electron transfer and reactivity, which can be examined via population-transfer dynamics as an experimentally interpretable surrogate observable.

Maintaining consistency across the entire reaction pathway often requires a large active space that integrates essential orbitals from different regions, and the computational cost of methods such as complete active-space self-consistent field (CASSCF) increases exponentially with the size of the active space. Consequently, even if the environment is hierarchically separated in a conceptually sound manner, the electronic structure problem of the reaction center itself remains a key practical limitation. Furthermore, the Born--Oppenheimer approximation~\cite{Born1927}, which allows the dynamic coupling between electronic and nuclear motions to be ignored, generally breaks down when electronic properties change dramatically along the reaction coordinate, further complicating the problem for classical computers~\cite{Domcke2011,Shu2023}. Such nonadiabatic and multistate effects are now widely recognized as central elements of reactivity and photochemistry in transition-metal systems and strongly correlated systems~\cite{Vogiatzis2019,Crespo-Otero2018}.

Quantum computing~\cite{Nielsen2010} has long been considered a promising approach for overcoming classical limits caused by the exponential growth of computational resources~\cite{Bauer2020,McArdle2020}. Aspuru-Guzik et al.\ demonstrated that quantum algorithms can suppress the scaling of key resource requirements to polynomial levels~\cite{Aspuru-Guzik2005,Wang2008}. In recent years, efforts to apply quantum computing to computational chemistry have made progress in various areas, including ground-state energy estimation, improved treatment of electronic correlations, and quantum simulation of reduced chemical Hamiltonians.

Research at the interface of quantum computing and chemistry has developed along two main directions, although the distinction between them is not always strict. Much of the work has focused on electronic structure problems. It typically relies on variational approaches or phase estimation to obtain ground- and excited-state energies. In practice, many studies have concentrated on relatively small molecular systems and on reducing measurement cost and circuit overhead. Another direction explores the simulation of model Hamiltonians and dynamical processes. In this context, time evolution is a natural target for quantum devices~\cite{Wang2023}. To make such problems tractable on current hardware, simplified descriptions are often introduced, for example active-space models or effective diabatic Hamiltonians.

With the advent of the NISQ (noisy intermediate-scale quantum) era, the focus has shifted from asymptotic advantage toward what can be achieved under realistic constraints. A key question is how reliably chemically meaningful observables can be reproduced in the presence of noise and limited circuit depth. Despite ongoing progress, it remains difficult to describe biomolecular reaction centers with high multireference accuracy across extended reaction coordinates. This has motivated the development of practical workflows for validating reduced Hamiltonians and their couplings.

In this work, we introduce a quantum--HPC hybrid approach and demonstrate its feasibility on current quantum hardware. We focus on the implementation of multistate dynamics and on the diagnostic observables that can be obtained from it. Closer integration between classical workflows and quantum implementations remains an important direction. The workflow proceeds in four stages. First, an effective multistate Hamiltonian is constructed along the reaction coordinate using classical electronic-structure calculations. Next, its spectral properties are verified. The Hamiltonian is then used to perform quantum time evolution and to obtain the dynamics of state populations from a reactant initial state. Finally, the population of the product manifold is analyzed as a diagnostic observable to identify near-degeneracy regions and possible deficiencies in the reduced model.

Unlike energy-based metrics, which are often insensitive to localized near-degeneracy, population-transfer observables depend explicitly on interstate couplings through the time evolution of electronic amplitudes. In standard nonadiabatic dynamics, the time evolution of state populations is described in terms of transition or hopping probabilities governed by nonadiabatic couplings~\cite{Tully1990}, and in perturbative electron-transfer regimes by rates that scale with the square of the electronic coupling~\cite{Blumberger2015}. Consequently, missing states or couplings can lead to significant deviations in population dynamics, providing a more stringent and chemically intuitive validation of reduced Hamiltonians.

We validate the dynamics using Quantinuum’s System Model H1 trapped-ion quantum computer ``Reimei,'' examining whether chemically interpretable multistate population transfer can be reproduced under realistic NISQ conditions. Here, ``chemically interpretable'' refers to linking population transfer with electronic-state interactions and state mixing, and with measurable observables such as spectroscopic signals. The platform provides all-to-all connectivity and supports mid-circuit measurement and conditional branching, which makes it suitable for such benchmarks.

This work makes two contributions. First, we present a dynamics-based diagnostic framework for assessing reduced Hamiltonians along reaction coordinates. Second, we demonstrate this framework end-to-end on trapped-ion hardware, including a coordinate sweep that identifies the region of strongest state mixing. To ensure hardware feasibility, we employ a one-hot mapping in the single-excitation subspace, enabling direct simulation of reaction-center dynamics from HPC-derived Hamiltonians. This allows practical validation of reduced models using dynamical observables on real hardware.

\section{Method\label{sec:method}}

We outline the workflow used to evaluate the validity of the reduced Hamiltonian and its implementation under NISQ setup. The workflow proceeds as follows. First, we describe how the reduced Hamiltonian is constructed, how the reference dynamics are obtained from classical time evolution, and how the model is mapped onto quantum circuits. We then define the population observable $p_P(t)$, and finally summarize the hardware and emulator setup, together with the analysis procedure used to separate different sources of error.

\subsection{Model system and physicochemical scope}

We focus on two key elements governing the cytochrome P450 reaction center during O$_2$ addition and redox processes: (i) electron transfer between the Fe 3d orbital and the O$_2$ $\pi^*$ orbital, and (ii) the resulting multistate mixing. We adopt the [Fe(CN)$_4$]$^{2-}$ complex as a simple structural analogue of the iron-porphyrin reaction center and treat [Fe(CN)$_4$(O$_2$)]$^{2-}$ formed by O$_2$ coordination as the core complex. The $\pi$-accepting CN$^-$ ligand promotes $\pi$ back-donation between the metal and ligand while suppressing excessive orbital delocalization, thereby facilitating the identification of Fe 3d orbitals and characterization of electron-transfer character toward O$_2$.

Since [Fe(CN)$_4$(O$_2$)]$^{2-}$ is a divalent anion, it is difficult to stabilize the Fe(III)--O$_2^{\bullet -}$ radical species (on the product side) as an isolated electronic state. Therefore, we introduced CH$_3$NH$_3^+$ as a counter cation and added four H$_2$O molecules as the minimal environment (Fe(CN)$_4$ complex + 4H$_2$O + CH$_3$NH$_3^+$). This setup allows us to concentrate the active space on Fe(3d) and O$_2$($\pi^*$) while keeping computational costs and structural diversity within manageable limits, enabling the observation of multistate potential energy curves and strong mixing regions along the selected reaction coordinate.

The spin multiplicity was fixed to a single value throughout the study. This is because diabatic coupling is non-zero only between electronic states with the same multiplicity. Therefore, a consistent quasi-diabatic Hamiltonian requires a predefined spin manifold. In order to obtain a more reliable energy criterion for selecting the working multiplicity, energy calculations for spin states were performed using a hybrid functional, as such methods are known to reasonably reproduce spin-state splitting in Fe complexes~\cite{Verma2017}. In this study, PBE0~\cite{Adamo1999} was employed because it provided stable SCF convergence and could be applied reliably in this system. To evaluate the relative stability of different spin states, geometric optimizations were performed separately for the triplet, quintet, and septet manifolds using a consistent set of basis functions (Fe: SBKJC (Stevens, Basch, Krauss, Jasien, and Cundari) effective core potential; O$_2$: 6-31+G(d); all other atoms: 6-31G(d))~\cite{Stevens1984,Stevens1992,Hehre1972,Hariharan1973,Clark1983}.

\begin{figure}
    \centering
    \includegraphics[width=0.99\hsize]{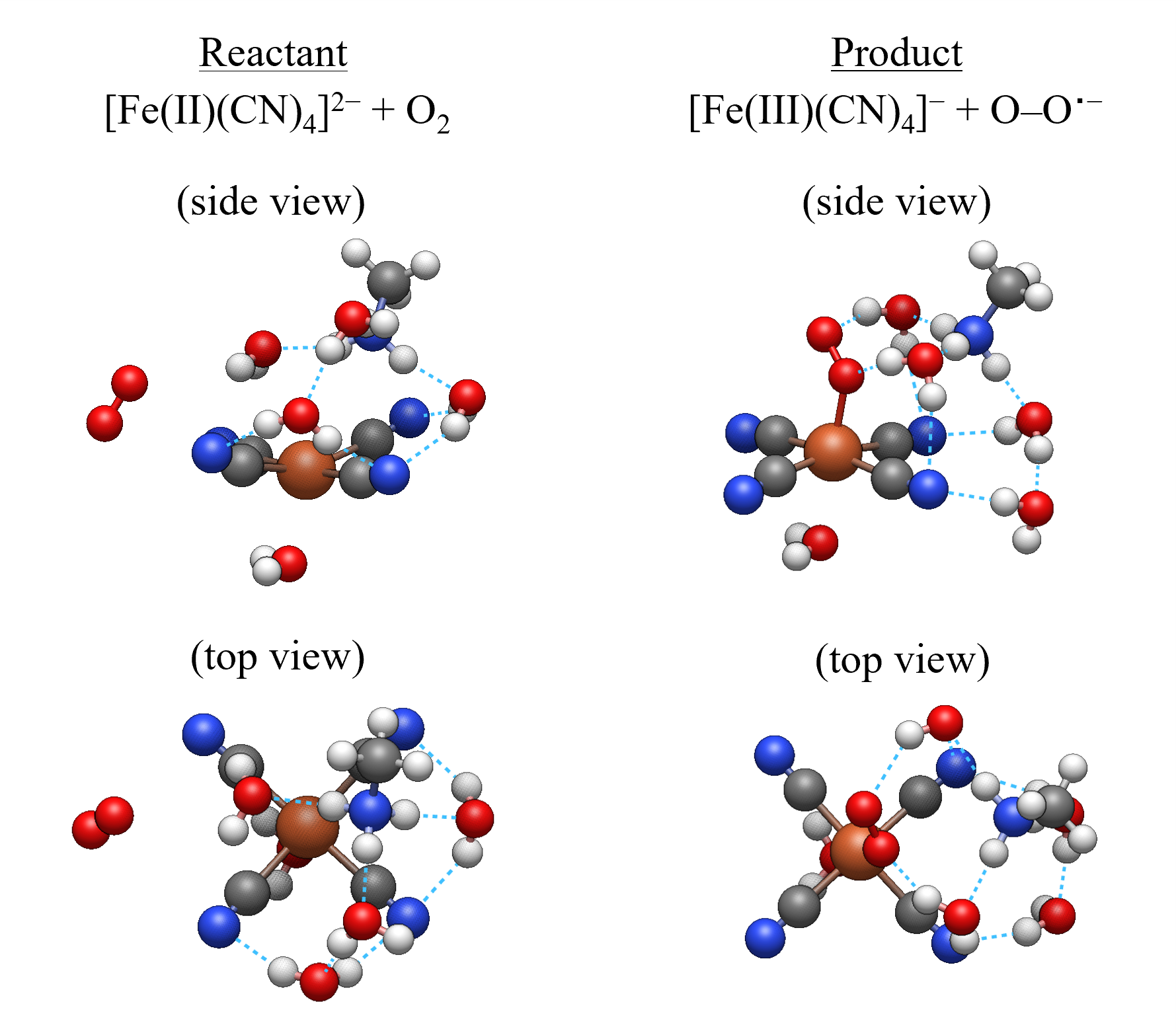}
    \caption{
        Molecular structures of the geometry-optimized reactant and product Fe complexes at the ROHF level.
    }
    \label{fig1}
\end{figure}

At the PBE0 level, the quintet is the lowest energy state, while the septet and triplet are located 0.08 eV (8.0 kJ mol$^{-1}$) and 0.36 eV (35 kJ mol$^{-1}$), respectively, in the [Fe(CN)$_4$(O$_2$)]$^{2-}$ model; therefore, the quintet state with a charge $-1$ was adopted as the physically reasonable reference multiplicity in all subsequent state-averaged CASSCF (SA-CASSCF) potentials, adiabatic/diabatic potential energy surface (PES) constructions, and electron-coupling calculations. The geometric structures and vibrational analyses of the endpoint states for the reactant (Fe(II)--O$_2$) and product (Fe(III)--O$_2^{\bullet -}$) were optimized at the restricted open-shell Hartree--Fock (ROHF) level using the aforementioned set of basis functions (Fig.~\ref{fig1}). All electronic-structure calculations were performed using GAMESS version 2024 R2~\cite{Schmidt1993,Gordon2005}.

\subsection{Reaction coordinate definition and structure generation\label{subsec:reaction}}

We introduce a dimensionless reaction coordinate $x$, where $x = 0$ corresponds to the reactant Fe(II)--O$_2$ structure and $x = 1$ to the product Fe(III)--O$_2^{\bullet -}$ structure. In practice, only the optimized endpoint geometries are available, so intermediate structures have to be generated to evaluate the multistate potential along the path. A straightforward interpolation, for example Cartesian interpolation or a simple distance scan, turns out to be problematic, due to the fact that it can easily lead to unrealistic distortions or even atomic clashes, especially when ligand rearrangement or solvent motion are involved. For this reason, such simple schemes are not reliable here.

To address this, we fix the two endpoint geometries as the optimized stable structures and generate 19 intermediate geometries between them using geodesic interpolation in internal coordinate space~\cite{Zhu2019}, giving a total of 21 geometries. The interpolation follows the method described in Ref.~\cite{Zhu2019} and is implemented with default settings in an available code base. This provides a geometrically continuous pathway without artificial distortions.

It should be noted that the purpose here is not to locate a minimum-energy pathway. Instead, we are mainly interested in testing the reduced Hamiltonian along a representative coordinate. For that reason, the intermediate structures ($0 < x < 1$) are left unrelaxed and treated as fixed points. This allows us to examine how electronic structure and interstate couplings change along the coordinate without additional optimization. To keep the orbitals continuity from one point to the next, we simply use the converged orbitals at one geometry as the initial guess for the following one.

\subsection{SA-CASSCF settings and active space}

SA-CASSCF calculations were performed at each geometry to evaluate the multistate electronic structure and state mixing in a consistent orbital basis~\cite{Roos1980,Werner1981,Ivanic2003}. Fifteen states were included with equal weights in the state average to avoid biasing orbital optimization toward a particular state and to mitigate root flipping in near-degeneracy regions.

\begin{figure}
    \centering
    \includegraphics[width=0.99\hsize]{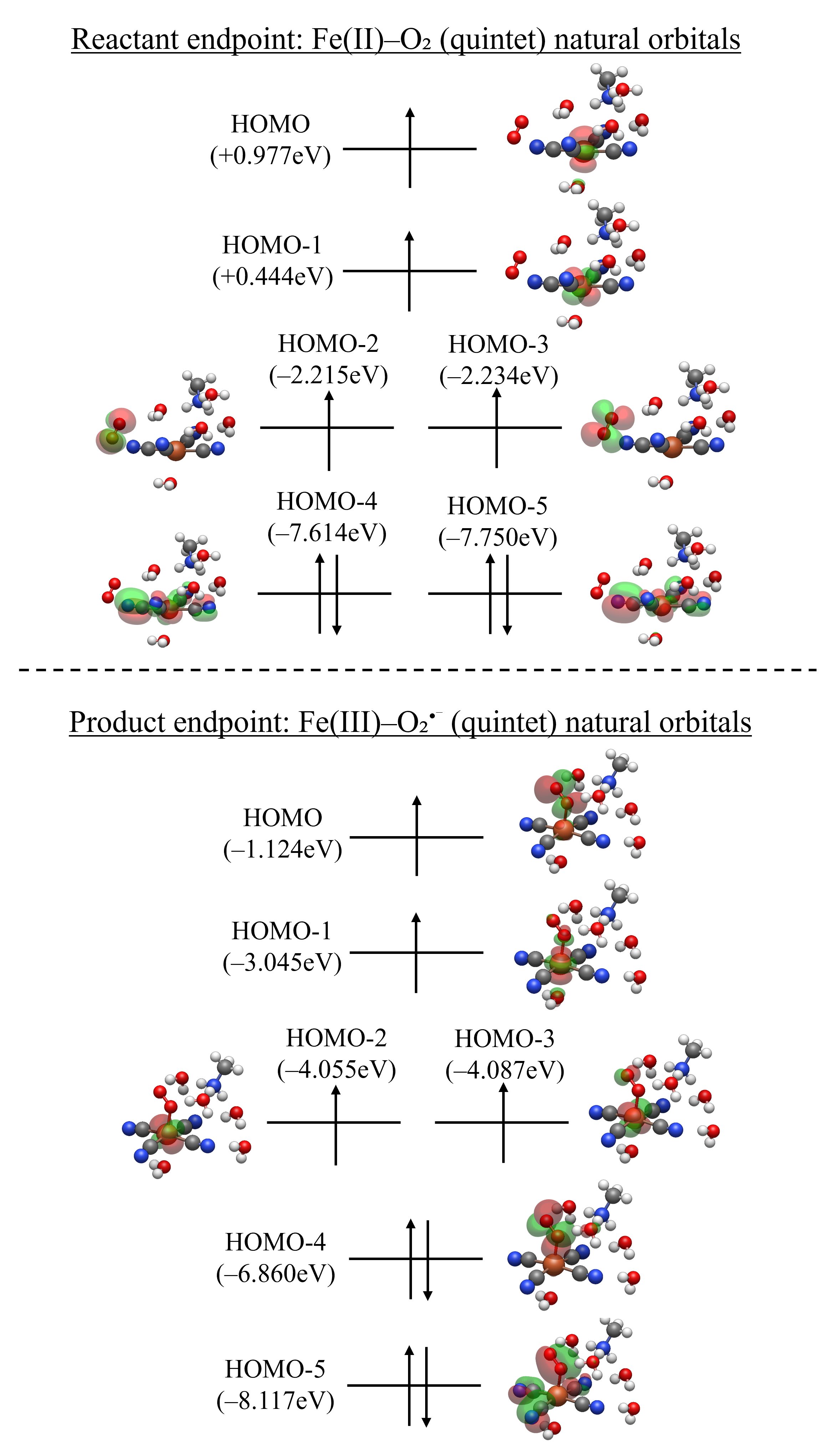}
    \caption{
        Active-space natural orbitals (HOMO--HOMO--5) and their orbital energies used to construct the SA-CASSCF (8e,6o) description at the Fe(II)--O$_2$ reactant and Fe(III)--O$_2^{\bullet -}$ product minima.
    }
    \label{fig2}
\end{figure}

The active space was fixed to CAS(8e,6o), including the Fe 3d orbitals and the O$_2$($\pi^*$) orbitals (Fig.~\ref{fig2}). This choice was designed to provide a physically justified minimal active space for reaction-center population-transfer analysis. Orbitals below HOMO$-5$ are largely localized on the CN ligands and act primarily as spectator orbitals; including them would introduce numerous ligand-centered excitations without materially improving the description of the Fe--O$_2$ electron-transfer dynamics targeted here. The selected active space therefore focuses the reduced model on the essential reaction-center degrees of freedom while maintaining multistate continuity along the reaction coordinate.

The distinct row table (DRT) parameters in the Graphical Unitary Group Approach (GUGA) module were set to IEXCIT = 2 (electron excitation level), NDOC = 2 (number of doubly occupied molecular orbitals in the reference), NALP = 4 (number of alpha spin singly occupied molecular orbitals in the reference), and NVAL = 0 (number of empty occupied molecular orbitals in the reference), corresponding to a configuration-interaction space including up to double excitations. The convergence thresholds were ACURCY = $1 \times 10^{-5}$ and ENGTOL = $1 \times 10^{-8}$.

To examine the consistency of the selected active space, we analyzed the CI expansions and the characters of the state-averaged natural orbitals. The orbital characters were estimated from the squared expansion coefficients of the natural orbitals. In the reactant, four orbitals are characterized by larger Fe 3d contributions ($\Sigma |C^2| \approx 1.2\text{--}1.5$) relative to O$_2$ p contributions ($\Sigma |C^2| \approx 0.2\text{--}0.5$), while two orbitals exhibit substantial O$_2$ p contributions ($\Sigma |C^2| \approx 1.0\text{--}1.2$). In the product, a similar distribution is found, with four orbitals dominated by Fe 3d contributions ($\Sigma |C^2| \approx 1.3\text{--}1.8$) and two orbitals showing enhanced O$_2$ p contributions ($\Sigma |C^2| \approx 1.0\text{--}1.1$). This confirms that the selected CAS(8e,6o) space spans the Fe 3d and O$_2$ $\pi^*$ manifold relevant to the metal--ligand charge-transfer configurations. The CI expansions show that the wavefunctions are dominated by a small number of configurations. In many cases, a leading configuration has a coefficient around 0.7--0.9, whereas in other states multiple configurations contribute with comparable weights (e.g., coefficients approximately 0.4--0.6). In all cases, several secondary configurations contribute with non-negligible coefficients (typically 0.1--0.3), indicating a consistent multiconfigurational character.

\subsection{State tracking and construction of the reduced Hamiltonian}

In the reduced model, the effective Hamiltonian is constructed using 14 energetically relevant states (State 1--State 14), while the highest state (State 15) is excluded. Along the reaction coordinate, State 15 remains well separated from the lower-energy states, and its mixing with them is negligible. For example, the energy gap between State 14 and State 15 is 30.3~eV at $x = 0$ and decreases to 11.9~eV at $x = 1$, but it remains sufficiently large to justify excluding this state from the dynamics.

At each point on the reaction coordinate, the state energies of States 1--14 were extracted from SA-CASSCF and used in the subsequent reduced-Hamiltonian construction and dynamics simulations. State tracking was performed to maintain chemically consistent labels along the reaction coordinate in the presence of possible state reordering near avoided crossings and near-degeneracy regions. In this way, each tracked label represents a one-to-one correspondence of electronic-state character across the full coordinate range.

\subsection{Fitting quasi-diabatic potentials and constructing the effective Hamiltonian}

Each quasi-diabatic potential $E_i(x)$ is represented as a sum of two Morse terms, a Gaussian term, and a constant offset. The parameters are fitted using least squares so that the resulting Hamiltonian reproduces the reference multistate energies (State 1--State 14) along the reaction coordinate $x$ (see Section~\ref{subsec:reaction})).
\begin{equation}
\begin{aligned}
E_{i}(x)
&= D_{e1,i} \left(1 - e^{-a_{1,i}(x - x_{e1,i})}\right)^2 \\
&\quad + D_{e2,i} \left(1 - e^{-a_{2,i}(x - x_{e2,i})}\right)^2 \\
&\quad + A_i \exp\left[-\left(\frac{x - x_{0,i}}{w_i}\right)^2\right]
+ \text{offset}.
\end{aligned}
\end{equation}
The fitting parameters corresponding to energy ($D_{e1,i}, D_{e2,i}, A_i$, and offset) are expressed in eV, while the remaining parameters ($a_{1,i}, a_{2,i}, x_{e1,i}, x_{e2,i}, x_{0,i}$, and $w_i$) are dimensionless to be consistent with the definition of the reaction coordinate $x$. The two Morse potentials capture the anharmonic behavior around the equilibrium region on the reactant and product sides, respectively, while the Gaussian term accounts for local structural barriers and rearrangement effects (e.g., solvent-shell rearrangement) along the reaction coordinate. To ensure numerical stability, the Morse exponents were clipped to avoid underflow/overflow, and a small positive offset was applied to $w_i$ to prevent division by zero in the Gaussian term.

Diabatic energy intersections were identified from sign changes in $E_i(x) - E_j(x)$ and used as indicators of regions where state mixing can occur. Using these intersections, we selected a subset of state pairs (``active pairs'') where the degrees of freedom of the parameters are concentrated in the subsequent coupling optimization.

The effective Hamiltonian was constructed as
\begin{equation}
H_{ij}(x) = E_i(x)\,\delta_{ij} + V_{ij}, \qquad V_{ii} = 0.
\end{equation}
where $\delta_{ij}$ is the Kronecker delta. The off-diagonal couplings $V_{ij}$ were determined by least squares such that the eigenvalues of $H(x)$ reproduce the reference multistate energies. Only the couplings for selected active pairs were optimized, while all other off-diagonal elements were set to zero. The resulting $14 \times 14$ Hamiltonian matrix is real-symmetric.

Here, ``quasi-diabatic'' denotes a smooth tracked-state representation constructed to preserve reaction-coordinate continuity rather than a unique \emph{ab initio} diabatization~\cite{Mandal2018,Grofe2017}. Because the off-diagonal couplings are inferred from eigenvalue matching, they are not strictly unique. Our aim is not to infer a unique diabatic model, but to build a sparse and physically motivated effective Hamiltonian whose adequacy can be interrogated through population-transfer dynamics. The effect of the coupling cutoff is evaluated by comparing the resulting population dynamics with the unpruned reference.

\subsection{Pruning: truncation of weak couplings\label{subsec:pruning}}

To reduce the circuit depth under hardware constraints, we truncated weak off-diagonal couplings using a threshold $\epsilon$ (in eV). Specifically, for the configuration $i \ne j$, we set $V_{ij}$ to zero in the case of $|V_{ij}| < \epsilon$, while leaving $E_i(x)$ unchanged. After truncation, we re-symmetrized $H$ to ensure numerical stability. To ensure a fair comparison, we consistently applied the same truncation rule to both the classical reference calculations and the quantum circuit generation.

\subsection{Classical reference dynamics: exact time evolution}

As a classical reference, the time evolution under the fixed reduced Hamiltonian $H$ at each coordinate was calculated exactly using an eigen-decomposition. With $H = V \,\mathrm{diag}(\lambda)\, V^{\dagger}$, the propagated state for any initial state $\psi(0)$ is given by
\begin{equation}
\psi(t) = V \,\mathrm{diag}\!\left(e^{-i \lambda t_{\mathrm{eff}}}\right) V^{\dagger} \psi(0).
\label{eq:time_evolution}
\end{equation}
Here, $t_{\mathrm{eff}}$ is the effective propagation time appearing in the phase factor and is defined as the physical time $t$ expressed in atomic units (a.u.). To ensure consistency, the same definition of $t_{\mathrm{eff}}$ was used in both the classical reference and the quantum circuit simulation.

\subsection{One-hot mapping and 14→12 reduction: effective subspace}

We adopt one-hot encoding, which maps an $N$-state system to a single-excitation subspace consisting of $N$ qubits. Here, the ground state $\ket{i}$ corresponds to a state in the computational basis where only the $i$-th qubit is excited. In this representation, off-diagonal hopping terms correspond to XY-type two-qubit interactions $\sigma_x^{(i)} \sigma_x^{(j)} + \sigma_y^{(i)} \sigma_y^{(j)}$.

In the truncated Hamiltonian used in this study, State13 and State14 are isolated from all other states (their off-diagonal couplings are zero), as they remain well separated from the other states and do not exhibit crossings along the reaction coordinate. Therefore, if the initial state has no amplitude in State13/14, the dynamics are confined to the remaining subspace. Consequently, by excluding State13/14, the hardware implementation is reduced to 12 states (12 qubits).

\subsection{Time-evolution circuit: Trotterization}

The time evolution operator $U(t) = e^{-iHt}$ is approximated using a first-order Trotter--Suzuki decomposition. The total time $T$ is divided into $M$ steps, defined as $\Delta t = T/M$. When $H$ is decomposed into implementable terms $\{H_k\}$, we implement
\begin{equation}
U(T) \approx \left[ \prod_k e^{-i H_k \Delta t} \right]^M.
\label{eq:trotter}
\end{equation}
Increasing $M$ reduces the digital (Trotter) error in the noiseless limit, but increases the depth of the circuit.

\subsection{Observable $p_{P}(t)$ and product-state set\label{subsec:observable}}

We define the product-side state set as $S_P = \{\text{State 1}, \text{State 2}, \cdots, \text{State 8}\}$ and evaluate the product-manifold population as
\begin{equation}
p_P(t) = \sum_{i \in S_P} \left| \braket{i \mid \psi(t)} \right|^2.
\label{eq:product_population}
\end{equation}
Here, $\ket{\psi(t)}$ is the time-evolved state under the reduced Hamiltonian. States 1--8 were classified as the product-side states because the potential stabilizes along the reaction coordinate toward the Fe(III)--O$_2^{\bullet -}$ side, allowing the product manifold to be identified as the low-energy region associated with the product character, which provides a physically intuitive definition for population analysis.

\subsection{Hardware/emulator conditions and post-processing\label{subsec:2-11}}

Hardware and emulator experiments were performed at 11 reaction-coordinate points, $x = 0.0, 0.1, \ldots, 1.0$ (a subset of the geodesically interpolated geometries). The propagation time was set to $t = 10\ \mathrm{fs}$ (413.4~a.u.), the coupling-truncation threshold was set to $\epsilon = 0.02$~eV (Section~\ref{subsec:pruning}), and the first-order Trotter step number was set to $M = 30$ for the hardware validation.

Two emulator settings were used for different purposes in this work. In Section~\ref{subsec:3-3}, the H1-Emulator was used to evaluate the generic accuracy--depth trade-off under a noise-aware circuit simulation. In Section~\ref{subsec:3-4}, the coordinate-resolved emulator analysis was performed under conditions matching the Reimei hardware, rather than a direct hardware comparison. Because these emulator settings may differ in calibration and compilation details, we discuss them separately.

The statistical sampling procedures differ between emulator and hardware measurements. In the emulator calculations, 10 independent batches of 4{,}000 shots were performed at each reaction coordinate point. The observable $p_P$ was calculated separately for each batch after postselection (Hamming weight = 1) and renormalization, and the final value was obtained as the average of the 10 batches. In contrast, the hardware experiment on ``Reimei'' was conducted using a single batch of 4{,}000 shots at each reaction coordinate point, reflecting practical constraints regarding device access and execution time. The proportion of events outside the single-excitation subspace was recorded as leakage.

The emulator results were obtained via statevector simulation using a noise model under the same Hamiltonian and time evolution settings. This model includes P2 (two-qubit gate) errors caused by imperfect entanglement operations and memory-related errors arising from decoherence during circuit execution and imperfections in state preparation and measurement.

\subsection{Reporting of statistical error, leakage, and postselection bias}

\begin{figure}
    \centering
    \includegraphics[width=0.99\hsize]{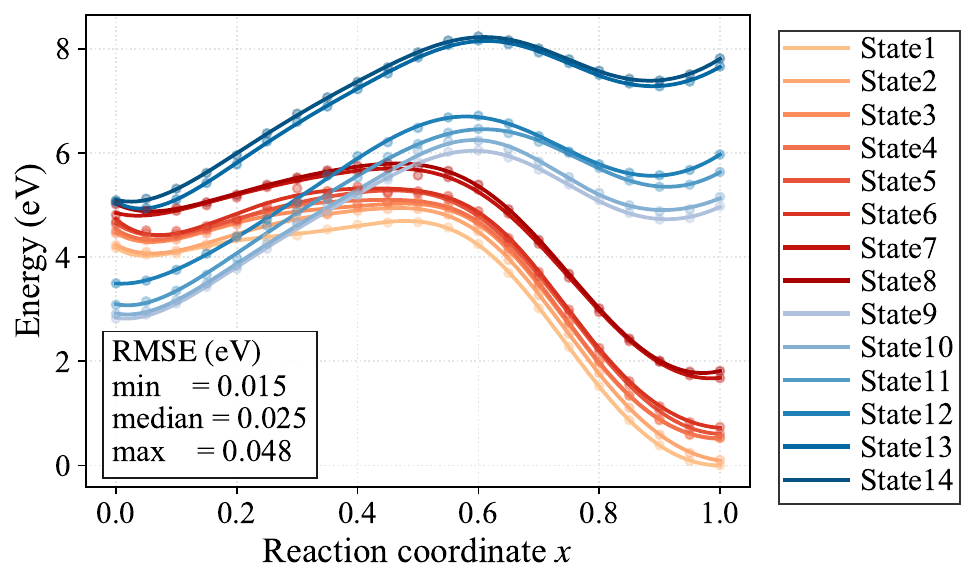}
    \caption{
        Quasi-diabatic potentials $E_{i}(x)$: reference points (markers) and fitted curves (lines). Warm colors denote product-side states (State 1--State 8) and cool colors denote reactant-side states (State 9--State 14).
    }
    \label{fig3}
\end{figure}

We report statistical uncertainties due to finite sampling (shot noise), leakage rates, and potential systematic biases introduced by postselection. This framework allows for the interpretation of results by separating approximation errors (e.g., Trotterization and truncation of coupling terms) from errors caused by hardware noise.

For emulator data, statistical uncertainty reflects both within-batch shot noise (finite sampling within each batch of 4{,}000 shots) and inter-batch variation across 10 independent runs, and it is quantified as the statistical spread across these batches. For hardware data, statistical uncertainty is evaluated from a single batch using a binomial approximation based on the effective postselection sample size $N_{\mathrm{eff}}$. It should be noted that the definitions of uncertainty estimation differ between emulator data and hardware data. While emulator uncertainty includes both sampling variation within a batch and inter-batch variation across independent runs, hardware uncertainty reflects only finite-shot noise within a single run.

\section{Results and Discussion\label{sec:results}}

In this section, we evaluate the validity of the reduced (effective) Hamiltonian constructed along the reaction coordinate and use the product-manifold population $p_P (t)$ as a diagnostic observable to quantify the region where electronic transitions are likely to occur. Given the constraints of NISQ hardware, we use the coupling cutoff threshold $\varepsilon$ and the number of first-order Trotter steps $M$ as control parameters to characterize the trade-off between accuracy and circuit resources (number of two-qubit gate count/depth). Finally, using the obtained operating conditions, we demonstrate end-to-end electronic dynamics on the trapped-ion quantum computer ``Reimei.''

 Before analyzing the dynamics, we verify the effective Hamiltonian at the spectral level. Since the reference energies are organized with chemically consistent labels through state tracking, we use spectral-level metrics for comparison with adiabatic eigenvalues (sorted by energy). Specifically, we fit $E_i (x)$ (quasi-diabatic potential) to a smooth functional form (Fig.~\ref{fig3}). The resulting potential exhibits a crossover between reactant-side and product-side states along the reaction coordinate, with product-side states becoming energetically more stable as $x$ approaches 1, which is consistent with a thermodynamically favorable reaction.

\begin{figure}
    \centering
    \includegraphics[width=0.99\hsize]{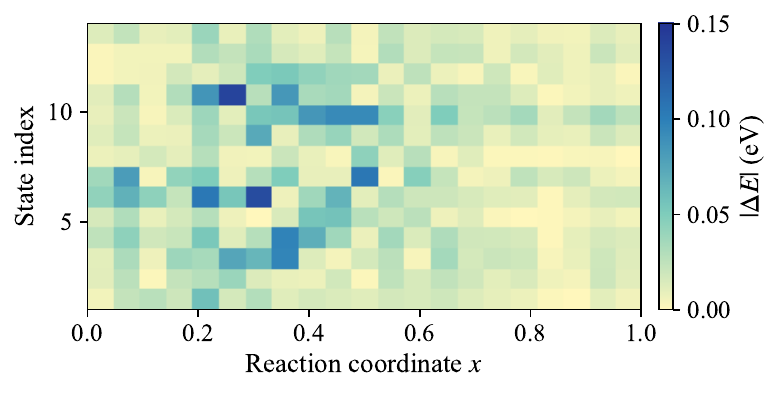}
    \caption{
        Hamiltonian validation by spectrum-level comparison between adiabatic eigenvalues of $H(x)$ and the SA-CASSCF reference energies, shown as absolute deviations $|\Delta E|$ after sorted eigenvalue matching at each $x$.
    }
    \label{fig4}
\end{figure}

Based on this, we optimized a sparse off-diagonal coupling $V_{ij}$ limited to the crossing pairs and confirmed that the adiabatic spectrum based on the Hamiltonian $H(x)$ reproduces the reference energies (Fig.~\ref{fig4}). The results evaluated at the spectral level (by comparing the sorted eigenvalues for each $x$) showed a consistency of $\mathrm{RMS} = 0.030~\mathrm{eV}$ and a maximum absolute deviation of $0.143~\mathrm{eV}$ across all $21$ points $\times$ $14$ states. This level of agreement ensures that the reduced Hamiltonian accurately captures the overall energy structure of the multistate manifold. Therefore, the reaction coordinate dependence of $p_P (t)$ can be interpreted as a feature of the Hamiltonian, specifically near-degeneracy and a small number of dominant interstate couplings, rather than a systematic error in the underlying energy surface. 

In Section~\ref{subsec:3-1}, we discuss the reaction coordinate dependence of $p_P (t)$ and its origin. In Section~\ref{subsec:3-2}, we evaluate how sparsification via $\varepsilon$ affects dynamics and circuit resources. In Section~\ref{subsec:3-3}, we summarize the precision-versus-resource curve for $M$ in the statevector (ideal)/emulator settings, in Section~\ref{subsec:3-4}, we analyze the emulator behavior in a coordinate-resolved manner, and in Section~\ref{subsec:3-5}, we discuss the reproducibility and limitations of the Reimei hardware by comparing it with the emulator at all points.

\subsection{Reaction-coordinate dependence of product-manifold population based on the effective Hamiltonian\label{subsec:3-1}}

Before discussing the reaction coordinate dependence of population-transfer dynamics, we clarify the scope of electronic-structure analysis used to interpret the origin of transfer enhancement. While electronic population analysis can, in principle, provide further insight into population-transfer mechanisms, due to the state-averaged nature of the density matrix, quantities such as electron population or charge distribution are not uniquely defined within the current SA-CASSCF framework. In particular, state-specific natural orbital occupancies and local spin state occupancies cannot be meaningfully extracted. Therefore, regarding the origin of the population-transfer enhancement near $x \approx 0.3$, we discuss it primarily from the perspectives of structural changes, the energetic proximity of electronic states, and their coupling, rather than using density-based indicators that may be ambiguously defined.

\begin{figure}
    \centering
    \includegraphics[width=0.99\hsize]{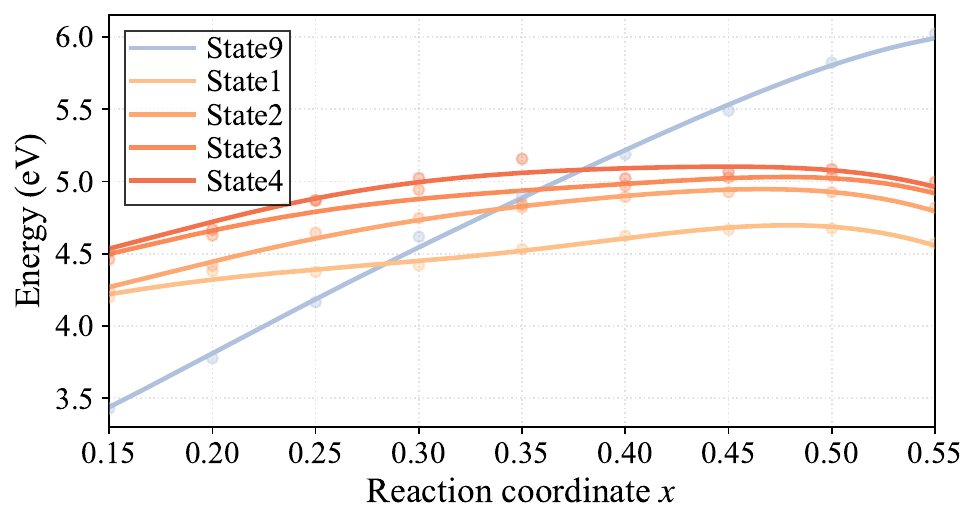}
    \caption{
        Key state energies (tracked labels) in the near-degeneracy region. Markers are SA-CASSCF reference points; solid lines are fitted $E_{i}(x)$.
    }
    \label{fig5}
\end{figure}

\begin{figure}
    \centering
    \includegraphics[width=0.99\hsize]{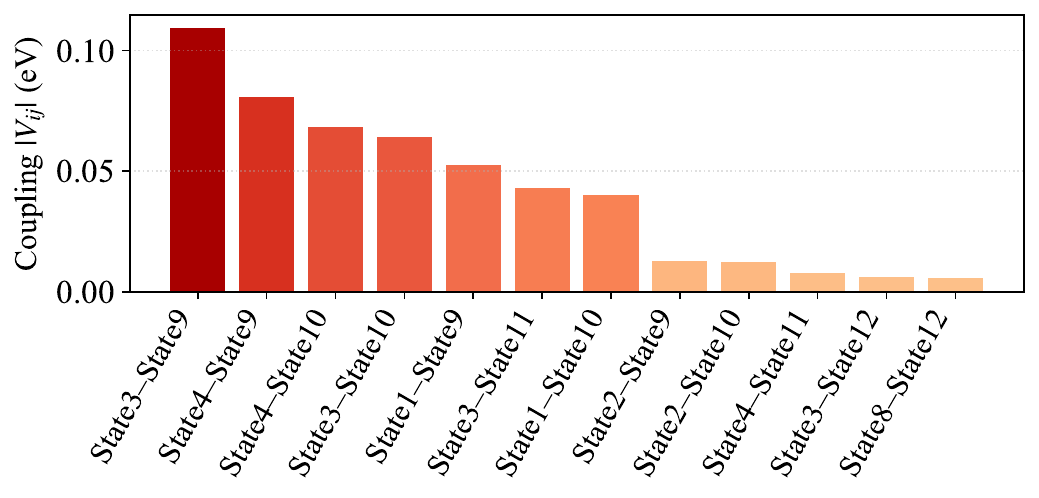}
    \caption{
        Dominant optimized couplings $V_{ij}$ ($\mathrm{eV}$) involving reactant-side states (State 9--12) and product-side states (State 1--8).
    }
    \label{fig6}
\end{figure}

Here, we use the $p_P (t)$ as a diagnostic observable to extract the transition sensitivity along the reaction coordinate. From the quasi-diabatic potentials $E_i (x)$, we find that a near-degeneracy occurs between the ground state of the reactant, State 9, and several product-side states in a narrow region around $x \approx 0.25$--$0.40$. As shown in Fig.~\ref{fig5}, this includes representative crossings between State 9 and lower product-side states (States 1--4), while similar quasi-degeneracies also occur for higher product-side states (States 5--8). In this region, there are a few relatively strong couplings, such as State 3--State 9 ($0.109~\mathrm{eV}$) and State 4--State 9 ($0.081~\mathrm{eV}$), which contribute to a Hamiltonian structure that promotes electronic transitions (Fig.~\ref{fig6}). Notably, the dominant couplings involve states with significant Fe 3d--O$_2$ $\pi^\ast$ character, as inferred from the active-space orbital composition (Fig.~\ref{fig2}) and the distribution of the largest optimized couplings (Fig.~\ref{fig6}), consistent with an electron-transfer-driven mixing mechanism in which metal-ligand backbonding plays a central role.

\begin{figure}
    \centering
    \includegraphics[width=0.99\hsize]{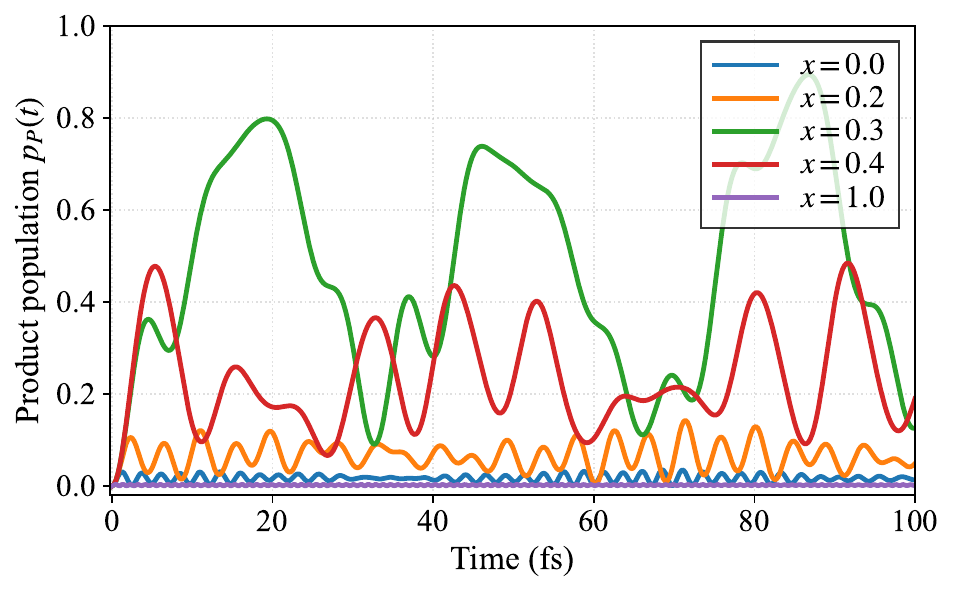}
    \caption{
        Exact time traces of the product-manifold population $p_{P}(t)$ under the reduced Hamiltonian $H(x)$, starting from the reactant-side ground state (State 9), at representative reaction-coordinate points $x = 0.0, 0.2, 0.3, 0.4,$ and $1.0$. The time evolution is shown up to $100~\mathrm{fs}$ to illustrate the robustness of the reaction-coordinate dependence beyond the initial population transfer.
    }
    \label{fig7}
\end{figure}

Next, we start from a one-hot initial state corresponding to the ground state of the reactants (here, State9) and evaluate the product-manifold population $p_P (t)$, as defined in Eq.~\eqref{eq:product_population}. Since the time dependence of $p_P (t)$ is sensitive to deficiencies in the reduced model (such as missing states or couplings), its use as a diagnostic indicator is justified. In the classical exact time evolution at a representative reaction coordinate point, the product-manifold population $p_P (t)$ shows a marked and rapid increase at $x = 0.3$, a more gradual increase at $x = 0.2$ and $x = 0.4$, and remains negligible at $x = 0.0$ and $x = 1.0$ (Fig.~\ref{fig7}). To evaluate the robustness of this trend, Fig.~\ref{fig7} shows the time evolution of the $p_P (t)$ extended to $100~\mathrm{fs}$.

At coordinates such as $x = 0.3$, $p_P (t)$ exhibits coherent oscillatory behavior with a characteristic period of approximately $30$--$35~\mathrm{fs}$. Since this timescale reflects the dominant electronic dynamics, the early-time regime ($t \lesssim 10~\mathrm{fs}$), corresponding to a fraction of the oscillation period, captures the initial buildup of population before a full oscillation develops. Consistently, the qualitative dependence on the reaction coordinate is already clearly established within the first $10~\mathrm{fs}$. That is, in contrast to the much weaker responses at $x = 0.0$ and $x = 1.0$, the strong collective accumulation of $p_P (t)$ at $x = 0.3$ becomes apparent well before the end of one oscillation period. Even when the propagation time is extended, the relative order of the $p_P (t)$ values along the reaction coordinate does not change.

To further support the choice of the observation time window, we analyzed the characteristic time scale of the population dynamics in the near-degeneracy region ($x \approx 0.3$). The Fourier transform of the time evolution of $p_P(t)$ revealed a dominant oscillation period of approximately $33$--$34~\mathrm{fs}$. This timescale quantitatively matches the electronic energy gap between strongly coupled states, and the characteristic period estimated from $T \approx 2\pi\hbar/\Delta E$ is approximately $34~\mathrm{fs}$. This indicates that the observed oscillatory behavior is dominated by coherent electronic mixing.

In contrast, vibrational analysis of the reactant and product structures showed that nuclear motions associated with the Fe--ligand framework occur on significantly longer time scales (typically $100$--$1{,}000~\mathrm{fs}$). This clear separation of the electronic and nuclear timescales supports the interpretation that the early time regime ($t \lesssim 10~\mathrm{fs}$) primarily reflects an electronically driven population transfer under effectively fixed nuclear geometries.

\begin{figure}
    \centering
    \includegraphics[width=0.99\hsize]{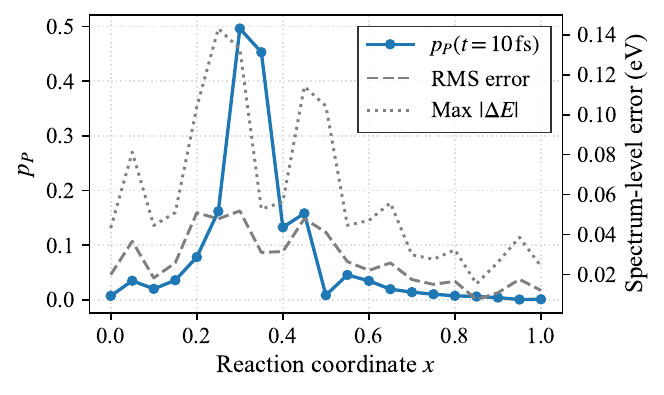}
    \caption{
        Reaction-coordinate dependence of the product-manifold population $p_P$ at $t = 10~\mathrm{fs}$. Markers denote the evaluated values. The pronounced maximum near $x \approx 0.3$ shows that the product-manifold population $p_P(x)$ selectively amplifies the coordinate region where near-degeneracy and dominant couplings most strongly enhance state mixing.
    }
    \label{fig8}
\end{figure}

Therefore, the value of $p_P$ evaluated at $t = 10~\mathrm{fs}$ serves as a representative early-time diagnostic observable that captures the intrinsic sensitivity of the reduced Hamiltonian to near-degenerate states and interstate coupling. Importantly, the qualitative dependence of $p_P$ on the reaction coordinate is robust within this early-time regime (i.e., within a fraction of the electronic oscillation period), indicating that selecting $t = 10~\mathrm{fs}$ does not introduce a bias in identifying regions where state mixing increases. Thus, in the following analysis, we use the $p_P (t = 10~\mathrm{fs})$ as a concise and physically meaningful summary measure of the transition tendency dependent on the reaction coordinate (Fig.~\ref{fig8}).

The quasi-diabatic potential shows that the intersection between the lowest reactant state (State 9) and the product states (States 1--8) is confined to a narrow window of the reaction coordinate. A representative crossing appears around $x \approx 0.25$--$0.40$. The resulting $p_P$ varies strongly with the reaction coordinate. It reaches a clear maximum near $x \approx 0.3$ (Fig.~\ref{fig8}), while remaining small near the reactant ($x \approx 0$) and product regions ($x \gtrsim 0.5$). This behavior follows the expected trend: transitions become more likely as the reactant and product potentials approach each other.

Although the effective Hamiltonian reproduces the adiabatic spectrum with small errors ($\mathrm{RMS} \approx 0.030~\mathrm{eV}$, maximum $|\Delta E| \approx 0.143~\mathrm{eV}$; Fig.~\ref{fig4}), these energy-based metrics show little structure as a function of $x$. In particular, no sharp feature appears near $x \approx 0.3$, where the population transfer is most pronounced.

This difference can be understood from the nature of the observables. Energy-based metrics average over the entire spectrum, whereas population dynamics are sensitive to couplings involving the initially occupied state and thus amplify local near-degeneracy effects. The observable $p_P(x)$ therefore displays a strongly inhomogeneous dependence on the reaction coordinate. It clearly identifies the near-degeneracy region where a small number of dominant couplings drive state mixing.

Taken together, these results indicate that reproducing the adiabatic spectrum alone is not sufficient to assess the validity of the reduced Hamiltonian. Population-transfer dynamics provide a more sensitive and chemically relevant diagnostic tool for evaluating the validity of the reduced models. In this system, the dynamics are governed primarily by a few strong couplings and the local level structure, rather than by many weak interactions. This observation motivates the following discussion on $\varepsilon$-based sparsification and the choice of the Trotter step number $M$.

\subsection{Balancing error and circuit-resources via coupling cutoff\label{subsec:3-2}}

To reduce two-qubit gate errors, we truncate weak interstate couplings and monitor the resulting dynamics using $p_P (t)$. A threshold $\varepsilon$ (in $\mathrm{eV}$) is introduced, and off-diagonal elements with $|V_{ij}| < \varepsilon$ are set to zero.

The optimized coupling matrix contains $32$ non-zero pairs. With the threshold $\varepsilon = 0.02~\mathrm{eV}$ (the value $\varepsilon = 0.02~\mathrm{eV}$ is specified in Section~\ref{subsec:2-11}), this number decreases to $7$ pairs (about $78\%$), which reduces the number of XY-type interactions per Trotter step in the one-hot representation. Only a small subset of couplings contributes significantly to the dynamics. The couplings exceeding $\varepsilon = 0.02~\mathrm{eV}$ correspond to a small set of dominant interaction pathways, namely State 3--State 9 ($0.109~\mathrm{eV}$), State 4--State 9 ($0.081~\mathrm{eV}$), State 4--State 10 ($0.068~\mathrm{eV}$), State 3--State 10 ($0.064~\mathrm{eV}$), State 1--State 9 ($0.052~\mathrm{eV}$), State 3--State 11 ($0.043~\mathrm{eV}$), and State 1--State 10 ($0.040~\mathrm{eV}$). These pathways remain after truncation.

\begin{figure}
    \centering
    \includegraphics[width=0.99\hsize]{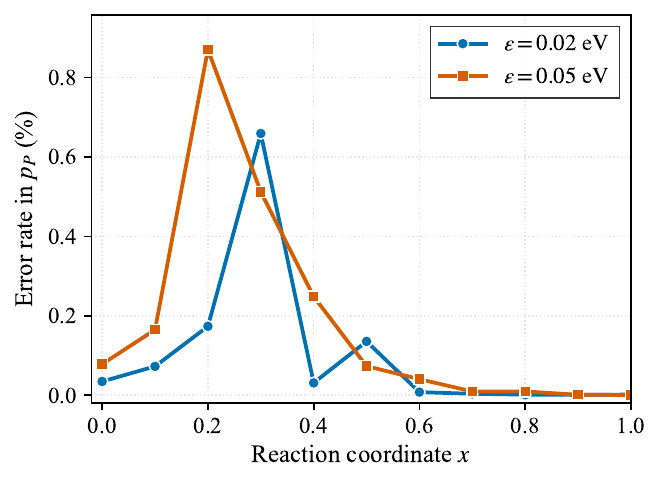}
    \caption{Effect of the coupling cutoff $\varepsilon$ on the error rate in the product-manifold population $p_P$ evaluated at $t = 10~\mathrm{fs}$. Symbols represent computed values at each reaction-coordinate point, and lines are guides to the eye. A cutoff of $\varepsilon = 0.02~\mathrm{eV}$ yields uniformly small errors across the reaction coordinate, whereas $\varepsilon = 0.05~\mathrm{eV}$ leads to larger and more widely distributed errors. The error rate (\%) is defined as $100 \times |p_P (\mathrm{approx}) - p_P (\mathrm{ref})| / p_P (\mathrm{ref})$, where $p_P (\mathrm{ref})$ is the corresponding classical reference value.}
    \label{fig9}
\end{figure}

We compare the effect of $\varepsilon$ using the relative error in $p_P$ against the classical time evolution under the same Hamiltonian. For $\varepsilon = 0.02~\mathrm{eV}$, the error stays small across the reaction coordinate, including the region near $x = 0.3$ where $p_P$ is largest (Fig.~\ref{fig9}). At $\varepsilon = 0.05~\mathrm{eV}$, the error becomes strongly coordinate-dependent and exceeds $0.2\%$ at several points, particularly near regions of strong coupling and near-degeneracy. The error also varies significantly with $x$ in this case.

Based on these results, we adopt $\varepsilon = 0.02~\mathrm{eV}$ as a practical threshold that preserves the dynamics while reducing circuit cost. This value is close to the thermal energy scale under ambient conditions ($k_{B}T \approx 0.025~\mathrm{eV}$), providing a useful reference.

\subsection{Accuracy-depth trade-off versus the Trotter step number: statevector/emulator\label{subsec:3-3}}

\begin{figure}
    \centering
    \includegraphics[width=0.99\hsize]{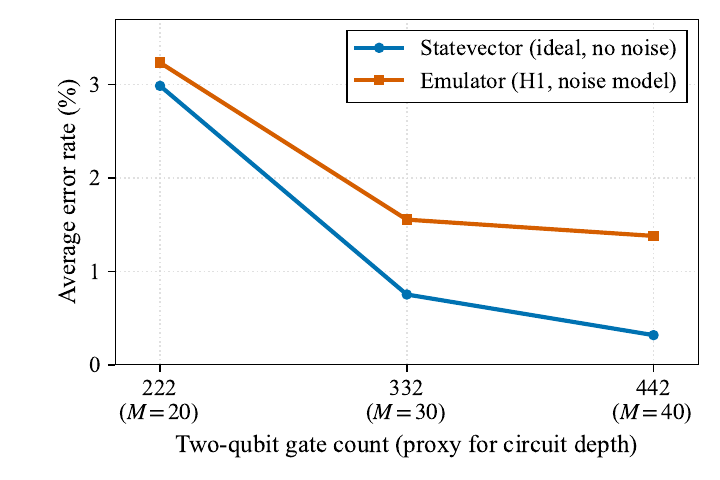}
    \caption{Accuracy-resource trade-off for first-order Trotterization at $t = 10~\mathrm{fs}$ and $\varepsilon = 0.02~\mathrm{eV}$. Markers show evaluated mean absolute error rates over all reaction-coordinate points for each corresponding two-qubit gate count, with the Trotter step number $M$ indicated, and straight-line segments connecting them as guides to the eye. The error rate (\%) is defined as $100 \times |p_P (\mathrm{Trotter}) - p_P (\mathrm{ref}_{\varepsilon})| / p_P (\mathrm{ref}_{\varepsilon})$, where $p_P (\mathrm{ref}_{\varepsilon})$ is the classical exact reference under $\varepsilon = 0.02~\mathrm{eV}$. Average error rate is computed as the mean of the absolute error rate over all reaction-coordinate points $x = 0.0, 0.1, \ldots, 1.0$ at $t = 10~\mathrm{fs}$.}
    \label{fig10}
\end{figure}

Next, we examine how the number of steps $M$ in the first-order Trotter--Suzuki approximation in Eq.~\eqref{eq:trotter} controls both the digital approximation error and the circuit depth, as determined by the number of two-qubit gates. In the noiseless limit, increasing $M$ causes the Trotter error to decrease monotonically; however, in a noisy setting, errors due to depth accumulate, leading to the derivation of an optimal value for $M$. In the one-hot representation adopted here, Trotter step counts $M = 20, 30,\ \text{and}\ 40$ correspond to circuits containing $222$, $332$, and $442$ two-qubit gates, respectively, and in Fig.~\ref{fig10}, these are used as practical approximations for circuit depth.

Using the observable $p_P (t)$, we calculated the average for all reaction coordinate points $x = 0.0, 0.1, \ldots, 1.0$ at $t = 10~\mathrm{fs}$ for (i) ideal statevector simulation (no noise) and (ii) the noise model emulator, and evaluated the relative error compared to the exact classical time evolution under the same Hamiltonian (Fig.~\ref{fig10}).

In the statevector (ideal) simulation, the average absolute error at all reaction coordinate points decreases monotonically with increasing $M$ ($3.0\%$ at $M = 20$, $0.8\%$ at $M = 30$, and $0.3\%$ at $M = 40$). This reflects a systematic reduction in the digital (Trotter) error.

A similar trend is observed in the emulator, where the mean absolute error decreases with an increase in the number of Trotter steps ($3.2\%$ at $M = 20$, $1.6\%$ at $M = 30$, and $1.4\%$ at $M = 40$). This indicates that the reduction in digital error is maintained even in the presence of noise. However, the improvement from $M = 30$ to $M = 40$ is relatively small ($0.2\%$), indicating that increasing $M$ beyond $30$ yields only a marginal improvement relative to the circuit cost.

Therefore, Fig.~\ref{fig10} provides operational guidelines for selecting the number of Trotter steps, $M$, under realistic hardware constraints. Increasing $M$ improves accuracy, but this also increases circuit depth and hardware resource requirements. Consequently, hardware experiments should use an operating region that balances accuracy and implementability. Based on this trade-off, we adopt $M = 30$ as the practical operating point for this hardware experiment.

\subsection{Coordinate-resolved analysis of emulator accuracy and error structure\label{subsec:3-4}}

Fig.~\ref{fig10} provides guidelines for selecting $M$ based on the accuracy--depth trade-off, but it does not reveal how residual errors depend on the reaction coordinate or affect physically relevant observables. Because the reported errors are averaged over all points, they cannot distinguish between uniformly small deviations and localized discrepancies in dynamically sensitive regions, nor do they provide insight into the physical origin of the errors.

\begin{figure}
    \centering
    \includegraphics[width=0.99\hsize]{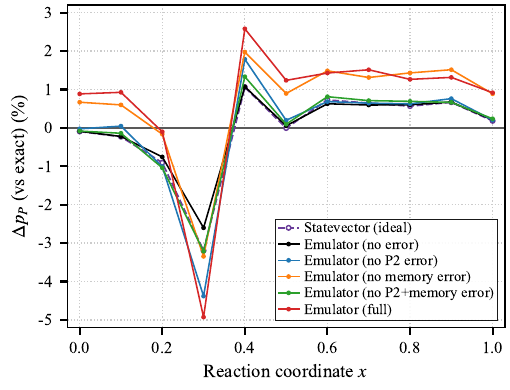}\\[2pt]
    \includegraphics[width=0.99\hsize]{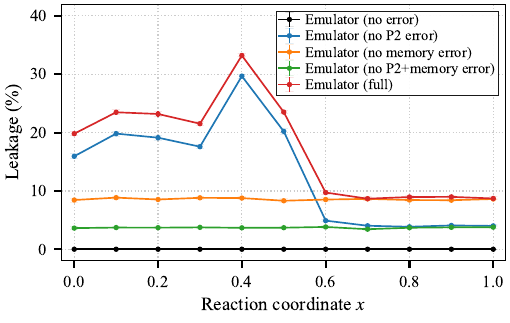}
    \caption{Verification of the emulator’s dynamics via reaction-coordinate decomposition under selected operating conditions ($t = 10~\mathrm{fs}$, $\varepsilon = 0.02~\mathrm{eV}$, $M = 30$). (a) Deviation of the product-manifold population from the classical exact time evolution, defined as $\Delta p_P (x) = 100 \times (p_P^{\mathrm{emu}} (x) - p_P^{\mathrm{exact}} (x))\ (\%)$. Curves are shown for multiple emulator configurations with different noise assumptions, including the ideal statevector case with no sampling or noise, a noiseless emulator, and an emulator variant with P2 and memory errors selectively removed. (b) Postselection leakage fraction (events outside the single-excitation subspace). This quantity provides a measure of the extent of out-of-subspace errors induced by the noise process and complements panel (a) by separating deviations arising within the computational subspace from those related to leakage.}
    \label{fig11}
\end{figure}

To address this limitation, we perform a verification of the coordinate-resolved population-transfer observable under the same operating conditions ($t = 10~\mathrm{fs}$, $\varepsilon = 0.02~\mathrm{eV}$, $M = 30$). At each reaction coordinate point, following the time evolution described by Eq.~\eqref{eq:time_evolution} from the initial state on the reactant side, we evaluate the population of the product manifold $p_P (t)$, as defined in Eq.~\eqref{eq:product_population}. For the emulator simulation, we calculate $p_P^{\mathrm{emu}}$ by summing the renormalized probabilities on $S_P$ from postselected measurement results restricted to the single-excitation subspace, i.e., Hamming weight $= 1$. The deviation from the classical reference value is $\Delta p_P (x)=100\times(p_P^{\mathrm{emu}} (x)-p_P^{\mathrm{exact}} (x))(\%)$, where $p_P^{\mathrm{exact}}$ is calculated from the classical exact time evolution under the same reduced Hamiltonian.

Concurrently, we quantify the postselection leakage as the proportion of measurement results located outside the single-excitation subspace. This quantity provides an orthogonal diagnostic measure for out-of-subspace errors induced by noise or circuit imperfections and is expected to be largely uncorrelated with the dynamical observables.

Fig.~\ref{fig11} shows both the $\Delta p_P (x)$ and leakage along the reaction coordinate. In panel (a), all curves are shown as deviations from the classical exact time evolution, allowing for a direct comparison of different error contributions across emulator configurations and statevector simulations. The $\Delta p_P$ panel directly measures the fidelity of the simulated dynamics, while the leakage panel (b) captures a dominant hardware-relevant error channel that may bias finite-shot estimates through postselection.

By comparing the two panels, it becomes clear whether the deviation in $p_P$ stems from inaccuracies in the simulated dynamics or from collective loss outside the intended subspace. This comparison allows for a more detailed interpretation of the sources of error by contrasting $\Delta p_P (x)$ with the leakage rate. As a useful perspective, the reaction coordinate can be divided into two regions: (i) the near-degeneracy window around $x \approx 0.25$--$0.40$ (where the dynamics are inherently sensitive), and (ii) the low-signal region ($x \lesssim 0.2$ and $x \gtrsim 0.5$), where the population transition is still small.

\begin{table*}[t]

\centering
\begin{tabular}{c|ccccc|ccccc}
\hline
$x$ &
\multicolumn{5}{c|}{Reimei} &
\multicolumn{5}{c}{Emulator} \\
\cline{2-6} \cline{7-11}
&
shots & $N_{\mathrm{eff}}$ & leakage (\%) & $p_{P}$ & CI95 &
shots & $N_{\mathrm{eff}}$ & leakage (\%) & $p_{P}$ & CI95 \\
\hline
0.0 & 4,000 & 3,128 & 21.8 & 0.007 & 0.0029 & $10\times4{,}000$ & 32,078 & 19.8 & 0.014 & 0.0013 \\
0.1 & 4,000 & 2,810 & 29.8 & 0.032 & 0.0065 & $10\times4{,}000$ & 30,611 & 23.5 & 0.026 & 0.0014 \\
0.2 & 4,000 & 2,833 & 29.2 & 0.058 & 0.0086 & $10\times4{,}000$ & 30,728 & 23.2 & 0.070 & 0.0028 \\
0.3 & 4,000 & 3,185 & 20.4 & 0.418 & 0.0171 & $10\times4{,}000$ & 31,393 & 21.5 & 0.432 & 0.0059 \\
0.4 & 4,000 & 3,057 & 23.6 & 0.188 & 0.0138 & $10\times4{,}000$ & 26,717 & 33.2 & 0.164 & 0.0053 \\
0.5 & 4,000 & 3,156 & 21.1 & 0.022 & 0.0051 & $10\times4{,}000$ & 30,597 & 23.5 & 0.021 & 0.0020 \\
0.6 & 4,000 & 3,322 & 17.0 & 0.039 & 0.0066 & $10\times4{,}000$ & 36,113 & 9.7 & 0.048 & 0.0019 \\
0.7 & 4,000 & 3,449 & 13.8 & 0.021 & 0.0048 & $10\times4{,}000$ & 36,535 & 8.7 & 0.028 & 0.0010 \\
0.8 & 4,000 & 3,542 & 11.5 & 0.012 & 0.0036 & $10\times4{,}000$ & 36,419 & 9.0 & 0.021 & 0.0016 \\
0.9 & 4,000 & 3,480 & 13.0 & 0.011 & 0.0034 & $10\times4{,}000$ & 36,409 & 9.0 & 0.016 & 0.0008 \\
1.0 & 4,000 & 3,343 & 16.4 & 0.013 & 0.0038 & $10\times4{,}000$ & 36,520 & 8.7 & 0.011 & 0.0011 \\
\hline
\end{tabular}

\caption{
Postselection statistics ($N_{\mathrm{eff}}$), leakage rates, and statistical uncertainties (CI$_{95}$), for both the hardware (Reimei) and the emulator runs. Here, $N_{\mathrm{eff}}$ is simply the number of shots that remain after we restrict to the single-excitation subspace, and CI$_{95}$ denotes the 95\% confidence interval. For hardware, we estimate CI$_{95}$ using a simple binomial approximation based on $N_{\mathrm{eff}}$: $\mathrm{CI}_{95} = \pm 1.96 \sqrt{p(1-p)/N_{\mathrm{eff}}}$. The emulator data are treated somewhat differently. Instead of a single batch, we run 10 independent batches (each with $4{,}000$ shots), and report the averaged values. The corresponding CI$_{95}$ is then estimated from the variation across these batches rather than from a binomial model.
}
\label{tab:table1}

\end{table*}

\begin{figure}
    \centering
    \includegraphics[width=0.99\hsize]{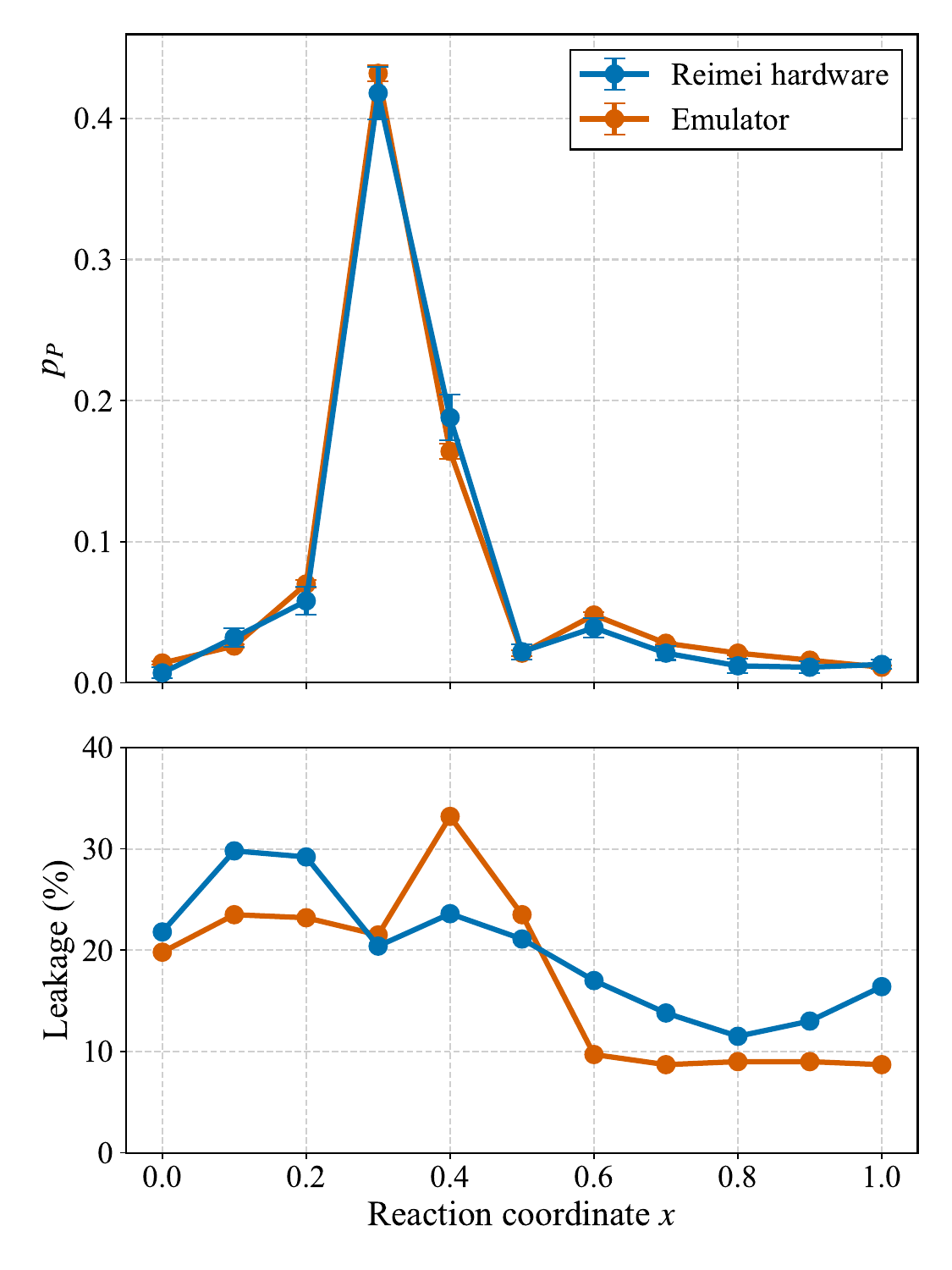}
    \caption{
        Hardware validation across reaction coordinate. (a) Product-manifold population $p_P$ measured on Reimei hardware ($4{,}000$ shots per point) and compared with the emulator ($10 \times 4{,}000$ shots per point; mean over 10 independent batches) at $t = 10~\mathrm{fs}$, $\varepsilon = 0.02~\mathrm{eV}$, and $M = 30$. Error bars indicate $95\%$ confidence intervals (hardware: binomial using $N_{\mathrm{eff}}$; emulator: across-batch CI). (b) Postselection leakage fraction (events outside the single-excitation subspace).
    }
    \label{fig12}
\end{figure}

In the near-degeneracy region, the product-manifold population is extremely sensitive to small perturbations of the effective Hamiltonian. Even in the absence of noise, near $x \approx 0.3$, the difference between classical exact evolution and the results from the statevector circuit---namely, the inherent digital (Trotter) error---becomes more pronounced. In this region, the emulator results show further deviations compared to the noiseless case, suggesting that noise processes such as P2 and memory-related errors further disrupt the coherent dynamics. This behavior reflects the fact that in regions where state mixing is strong, slight inaccuracies in coupling or phase are amplified into observable deviations in the $\Delta p_P (x)$.

In contrast, outside the near-degeneracy region, the underlying state $p_P$ is small, and the behavior is dominated by statistical and postselection effects, reducing the influence of coherent amplification. In these regions, leakage plays a more direct role by increasing statistical uncertainty through a reduction in the effective postselection sample size $N_{\mathrm{eff}}$. Table~\ref{tab:table1} shows that leakage exhibits a clear coordinate dependence (for example, leakage is high near $x \approx 0.1$--$0.2$ and decreases as $x$ approaches $0.7$). This suggests that the statistical reliability of postselection estimates varies depending on the coordinate.

However, the weak correlation between $\Delta p_P (x)$ and the leakage across the full reaction coordinate indicates that the deviation in $\Delta p_P (x)$ is not primarily determined by population loss outside the subspace. Furthermore, a significant portion of this discrepancy stems from perturbations within the single-excitation subspace, arising from changes in effective dynamics due to coherent gate errors and noise. 

Taken together, these results suggest that (i) the near-degeneracy region is primarily constrained by Hamiltonian sensitivity and intrinsic digital errors, (ii) leakage primarily affects statistical accuracy through $N_{\mathrm{eff}}$ and postselection, and (iii) the discrepancy between the emulator and the exact computation indicates the presence of noise mechanisms within the subspace in addition to leakage.

\subsection{End-to-end demonstration of reduced-Hamiltonian dynamics on quantum hardware\label{subsec:3-5}}

In this section, we perform end-to-end measurements of the product-manifold population $p_P$ on the trapped-ion quantum computer ``Reimei'', based on the operating conditions described above ($\varepsilon = 0.02~\mathrm{eV}$, $M = 30$, and $t = 10~\mathrm{fs}$).

Fig.~\ref{fig12} and Table~\ref{tab:table1} compare the results from the Reimei hardware and the emulator at 11 reaction coordinate points $x = 0.0, 0.1, \ldots, 1.0$. Both the hardware and the emulator reproduce the qualitative dependence on the reaction coordinate. The agreement with the noise-aware emulator demonstrates that the observed population transition is not a hardware-specific artifact but is due to the Hamiltonian structure. Specifically, they exhibit a pronounced maximum near $x = 0.3$ (hardware: $p_P \approx 0.418$; emulator: $p_P \approx 0.432$) and low-signal behavior (typically less than a few percent) at $x \gtrsim 0.5$.

Although quantitative deviations are observed at some reaction coordinate points, the hardware results consistently reproduce the reaction coordinate dependence of $p_P$ qualitatively, including a prominent maximum near $x \approx 0.3$ and low-signal behavior away from the near-degeneracy region. This confirms that the end-to-end workflow generates chemically interpretable dynamics on actual hardware under realistic NISQ conditions.

Small deviations between hardware and emulator are primarily attributed to coherent errors from Trotterization and hardware noise. Differences in leakage are attributed to imperfections in the noise model and hardware-specific error channels, but they show only weak correlation with population deviations and do not affect the qualitative conclusions. 

Taking the above analysis as a whole, this end-to-end workflow has demonstrated that chemically interpretable multistate electronic dynamics can be obtained on real quantum hardware under realistic NISQ constraints, based on a reduced reaction-center Hamiltonian constructed from classical multireference calculations. The current implementation, however, is limited to fixed-nucleus dynamics within a compact effective subspace and does not yet account for nuclear motion or environmental fluctuations. Building on this demonstration, future work will extend the current framework to larger effective Hamiltonians and longer-time-scale dynamics. It will also incorporate nuclear motion and environmental fluctuations into dynamics-based verification.

\section{Conclusions and outlook\label{sec:conclusions}}

A central challenge in mechanistic modeling of transition-metal reaction centers is not only the construction of reduced active-space Hamiltonians, but also the assessment of whether the retained subspace and interstate couplings remain dynamically adequate along a reaction coordinate. In this work, we addressed this problem by using product-population dynamics as a diagnostic observable for active-space-derived reduced Hamiltonians.

Starting from multistate SA-CASSCF data along the reaction coordinate, we assembled a reaction-coordinate-dependent effective Hamiltonian, validated it at the spectrum level ($\mathrm{RMS} \approx 0.030~\mathrm{eV}$; maximum deviation $\approx 0.143~\mathrm{eV}$), and showed that the product-manifold population $p_P (t)$ identifies a narrow near-degeneracy window around $x \approx 0.3$ where state mixing is most pronounced. By combining coupling pruning ($\varepsilon = 0.02~\mathrm{eV}$) with first-order Trotterization ($M = 30$), we defined a practical hardware-executable operating regime and reproduced the key coordinate-dependent population-transfer trend on the trapped-ion quantum computer Reimei.

These results establish a dynamics-based diagnostic route for assessing reduced-Hamiltonian adequacy on current quantum hardware. At the same time, the present study concerns electronic dynamics at fixed nuclear geometries along an interpolated reaction coordinate and therefore does not constitute a full reactive molecular-dynamics treatment. Extending the workflow to include nuclear motion, environmental fluctuations, refined emulator calibration, and more adaptive reduced-model construction remains an important next step.

\section*{Data availability}

Additional data are available from the corresponding authors upon reasonable request. 

\section*{Acknowledgments}

We are grateful to the entire Quantinuum team for their many contributions to this study. We thank Miori Hiraiwa for insightful discussions that helped shape the early stages of this work; Nathan Lysne, Craig Holliman, Tomohiro Murata, and Yuga Kodama for their support in running the circuits on Reimei; Nobuhiko Kido for his invaluable support in project coordination and management; and Cono di Paola, Kesha Sorathia, and Tyler Leblond for useful discussions and feedback on the manuscript. Part of this study is based on results obtained from a project, JPNP20017, commissioned by the New Energy and Industrial Technology Development Organization (NEDO). This work was also supported in part by the COE research grant “Quantum-HPC Hybrid Application Hub” in computational science from Hyogo Prefecture and Kobe City through the Foundation for Computational Science (FOCUS), and was conducted as part of the project “Development of Drug Discovery AI Platform through Industry–Academia Collaboration-DX (DAIIA-X),” supported by the Japan Agency for Medical Research and Development (AMED) under Grant Number JP26nk0101112. 

\bibliography{main}

\end{document}